# The market drives ETFs or ETFs the market: causality without Granger

Peter Lerner[1]

Abstract

This paper develops a deep learning-based econometric methodology to determine the causality of the financial time series. This method is applied to the imbalances in daily transactions in individual stocks, as well as the ETFs reported to SEC with a nanosecond time stamp. Based on our method, we conclude that transaction imbalances of ETFs alone are more informative than the transaction imbalances in the entire market. Characteristically, a sheer number of imbalance messages related to the individual stocks dominates the imbalance messages due to the ETF in the proportion of 8:1.

**Introduction**

Since the 2010s, the transactions in the stock market began to bear nanosecond time stamps. This required new methods of analysis adapted to the new realities (Hasbrouck, 2010). Yet, the capacity of the human mind to analyze multidimensional time series consisting of billions of market events remained unchanged. Because of our evolution in three-dimensional space, humans have the best grasp of two-dimensional information. Consequently, the image analysis methods are among the best developed in the whole of signal processing.

My paper adapts deep learning methods developed for image processing. The use of C-GAN (Convolutional Generational Adversarial Neural Networks) has been adopted since 2015 in the image analysis/generation of fake images (Goodfellow, 2017). The essence of the C-GAN is that the network is divided into two parts: generator and discriminator/critic (Rivas, 2020). The generator net produces fake images based on noise and learns to make them better. The discriminator tries to distinguish fake from real images based on training statistics.

To adapt this method to the financial time series, I use preprocessing methods to transform financial time series into two-dimensional images of realistic size. These images are standardized to ensure the comparability of the procedure. I use the method to analyze trading imbalances, required to be stored by the SEC. The imbalance events constitute the situation when the counterparty would not deliver the stocks in the amount it transacted. The number of these events per day is several million.

Why this problem is important? Given the instances of "flash crashes" in the market, the first and the largest of the reported being the Flash Crash of 2010 on NYSE, the role of the ETF took the forefront of the financial research. In particular, the Flash Crash on NYSE was attributed to the behavior of liquidity of the S&P minis (Easely, 2013). Because of the explosive growth in the ETF markets (more information in Section 1), the traded volume in the secondary assets (ETF shares) can easily exceed the volume of the underlying securities: stocks or futures. Intuitively, this can cause a problem in the delivery of the underlying, which propagates through the system and causes an interruption in the orderly work of the market.

---

[1] Anglo-American University, ul. Letnikow, 120, Prague. pblerner18@gmail.com, pblerner@syr.edu

In the current paper, I explore whether the market imbalance events drive ETF imbalances or the other way around. This problem seems to be a good testbed for the proposed deep learning method.

The structure of the paper is as follows. In the first section, I briefly outline commonly referred information on the ETF market. In the second section, I describe the structure of the database. In the third section, preprocessing of the SEC data into two-dimensional images is described. In the fourth section, we establish the causality between the ETF transaction imbalances and market transaction imbalances. In conclusion, we summarize our results.

## 1. Formulation of the problem and the literature review

The ETF markets exhibited explosive growth in recent years ( (NYSE, 2021)). Invented as a way to securitize stock inventories of the large asset managers but developed into a major asset class in their own right (Gastineau, 2010). Different anomalies in the markets, in particular the Flash Crash in May 2010, were partly attributed to the instability of the exchange-traded products (ETPs), especially the S&P minis (Easely, 2013).

The question of whether the introduction of ETFs increased, or decreased market quality has been discussed in many authoritative publications (Bhattacharya A., 2018), M. O'Hara, and *op.cit.*). I. Doron *et al.* (Doron, 2017) mentioned that there are two main opposing reasons for expectations of market quality change. On the positive side, exposure to the ETF provides more information on the underlying stocks, in particular, less liquid stocks with little analyst coverage. On the negative side, the uninformed traders tend to exit the market in the underlying securities in favor of ETFs thus depressing liquidity. Furthermore, much of the ETF activity happens at the market close (Nolan, 2016). Nolan *et al*. noticed that ETFs typically have larger mispricing and wider spreads during end-of-trading, especially on the days of most volatile trading.

If the influence of the ETFs is so prominent, can they be a catalyst for the extreme events in the market? Some authors, e.g., A. Madhavan answer in the positive (Madhavan A., 2019). At least, there is a recognition that there are new kinds of risks inherent in the proliferation of the ETF (Pagano, 2019). If that is true, can big data analyses and deep learning instruments provide some warning whether the extreme events can be coming?

To begin answering this question, we develop a structured methodology, which allows us to determine with some certainty whether ETF illiquidity is the result of market fluctuations or it is the other way around? The hypothetical mechanism is as follows: ETF trading initiates the delivery of ETF constituents ("in-kind" transaction) or a cash equivalent ("in-cash" transaction) if the underlying assets are easily tradable (Bhattacharya A., 2018). If the aggregate volume of ETF executions were small and/or evenly spread in time, this would introduce frictions in orderly market execution.

And indeed, there are few inherent inefficiencies. First, there is no clarity as to whether trade interruption results in actual economic events in one or more underlying stocks or from the effects of stock aggregation by the ETFs, for instance, changes in the credit risk of the ETF swap counterparty.

Second, because ETF transactions are prominent in hedges, they are highly non-uniform throughout the day (Nolan, 2016). The company, which missed the delivery can wait till the end of the day to close the deal when the market will be more stable. And this list can be continued. This paper does not judge whether the ETFs are a "good" or "bad" influence on market liquidity. It strives to clarify the enormous influence ETFs have on the stock market, in particular, the direction of information transmission.

According to the foundational models of market microstructure, the principal driver of the price changes is the imbalances in supply and demand for given security (Kyle, 1985), (Glosten A. R., 1985). At the current time, imbalance messages can be followed with nanosecond precision. There is probably little value added to further increasing precision because signals cannot propagate but a few meters—i.e. the size of the trading room—with already achievable latency (Bartlett III, 2019). One of the first studies going up to the nanosecond granularity was (Hasbrouck, Price discovery in high resolution, 2021). This wealth of the available data creates many problems of its own. The human mind is poorly adapted to rationalize such amounts of data. Furthermore, our senses evolved in 3D space and have difficulty comprehending multidimensional datasets.

One of the principal channels of influence of the ETF market on the overall market stability is using the ETF shares for shorting and hedging. ETF shares are "fail-to-deliver" on settlement for different reasons. This could be actual economic troubles of the company using ETFs, shorting in expectation of a real price movement by the AP (Authorized Provider), the analog of the Market Makers (MM) for stock, or "operational shorting" (Evans, 2018). The description of the operational shorting in the above-cited paper by Evans, Moussawi, Pagano, and Sedunov is so exhaustive that I provide a somewhat long quote from that paper.[2] Before a security is classified as "fail-to-deliver", the imbalance record is created. Usually, the imbalance is cleared by the end-of-day trading, or the next day before the trading hours. The reputation penalty of being cited in imbalances is usually small (Evans, 2018).

This work is dedicated to the research of the methods, which can rationalize imbalance datasets with nanosecond time stamps. We compress them into two-dimensional "fingerprints", for which a rich array of algorithms developed for the analysis of the visual images is already available. The dataset we use is the list of imbalance messages provided by NYSE Arca. "NYSE Arca is the world-leading ETF exchange in terms of volumes and listings. In Nov. 2021, the exchange had a commanding 17.3% of the ETF market share in the US" (investopedia.com, 2021). The special significance of the data for our problem setting is illustrated by the fact that a glitch in the NYSE Arca trading system influenced hundreds of ETF funds in March 2017 (NYSE Arca suffers glitch during auction, 2017).

---

[2] "One important objective of APs in the primary ETF market is to harvest the difference between ETF market price and its NAV… As demand for the ETF grows from investors in the secondary market, the ETF's market price should increase [Increasing the possibility of market arbitrage—P. L.] However… selling ETF shares and buying the underlying basket/creating the ETF shares are not necessarily instantaneous. The AP sells the new ETF shares to satisfy a bullish order imbalances but can opt to delay the physical share creation until a future date. By selling ETF shares that have not yet been created, the AP incurs a short position for operational reasons… that we hereafter call an "operational short" position." The paper (Evans, 2018) also lists "directional shorting", i.e. speculation on the changing market price as a reason for "fail-to-deliver".

Messages in our database have the following types: type '3', type '34', and type '105'. The message type 3 is a symbol index mapping (reset) message. The message 34 is a security status message, which can indicate 'opening delay', 'trading halt' and 'no open/no resume status. Finally, the message type 105 is an imbalance message. More information about format and content of the messages and the datasets can be found in Appendix and (TAQ NYSE ARCA integrated feed client specification, 2014)

The number of imbalance messages (type 105) for each trading day is around four million and each message consists of 15-20 standardized fields. TAQ NYSE Arca equities—TAQ NYSE imbalance files provide "buy and sell imbalances sent at specified intervals during auctions throughout the trading day for all listed securities" (TAQ NYSE ARCA integrated feed client specification, 2014).

## 2. Preprocessing – formation of the state variables database

We select the following variables: (1) number of messages per unit time, price, (2) dollar imbalance at the exception message, and the (3) remaining imbalance at settlement. The latter is rarely different from zero because a failure to rectify stock imbalances at the close of the trading session indicates a significant failure of market discipline and may entail legal consequences.

Our data can be divided into two unequal datasets: market messages in their totality and ETF-related messages, the first group encompasses the second. Because of the large volume of the data, we used an algorithmic selection of data into the ETF group. The messages in the datasets contain identifier 'E' for the Exchange Traded Funds, but in other places 'E' can indicate corporate bonds but it is too common a letter to filter for it in a text file. Instead, we chose separate data on market participants provided by NYSE, of which we filtered the names explicitly having the words "exchange traded" or a "fund". This identification is not 100% accurate because some closed-end mutual funds, which are not ETF could have entered our list, but they are expected to be dominated by ETFs.

We were left with 1061 names, which are automatically selected from the messages file. The number of daily events related to our list can be half a million or more so sorting by hand would be difficult, if not impossible.

We further group our data as follows: first, is the number of the type 105 messages per 100 seconds. Second is the cumulative imbalance per every 100 seconds in a 12½-hour trading day[3] in each out of the price bins.[4] The number of price bins is chosen approximately equal to the number of time intervals. Dollar imbalances are calculated by the following formula:

$$\$Imb_t = p \cdot (Imb_t - Settle_{400}) \qquad (1)$$

---

[3] Our messages begin at 3:30 AM and end at 4 PM, 45,000 seconds in total, usually, but not always, 449 100-second intervals beginning with zero. Each time interval contains ~9,000 messages on average. Yet, the highest message rate can be more than twenty times as high.

[4] The price bin methodology is reminiscent of the VPIN measure of Easely, O'Hara, and Del Prado (2013). We experimented with linear as well as the logarithmic scale of our data. In this paper, we use a logarithmic scale.

Where $p$ is the last market price, $Imb_t$ is the undelivered number of shares, and $Settle_{400}$ –the number of shares unsettled by the end of the trading session, usually at 4:00 PM.

The 100-second intervals were chosen arbitrarily but with the intent to have a two-dimensional data tensor being processed on a laptop and having sufficiently nice statistics. The imbalances are distributed quite irregularly around the 45,000 second period and can be visually grouped into the "beginning of the day settlement", "midday settlement" and "end-of-day settlement" (see Fig. 1).

[Place Figure 1 approximately here]

As expected, most of the dollar imbalances are small. To avoid data being swamped into a trivial distribution—a gigantic zeroth bin—and a uniformly small right tail, we use a logarithmic transformation for all variables including time: $\tilde{x}_t = \ln(1 + x_t)$. Unity is added to deal with the zeroes in the database.

The plot of the summary statistics for a typical day is shown in Fig. 2. We observe that the maximum number of imbalance messages created by ETFs for the 100 second period during a trading day is about one-eighth of the total number of exceptions (~25,000:200,000) in the market but the cumulative value of imbalances created by the ETFs is about 60% of the total. This suggests that average imbalances are much higher when the ETF shares are involved.

[Place Figure 2 approximately here]

The final step is making our data amenable for the deep learning algorithms, which were mostly designed to deal with visual data ("fake images"). We further compress our non-uniform rectangular matrices – in some cases, messaging did not begin at exactly 3:30 AM, etc. One data file included only NYSE into 96×96 squares, which we call "fingerprints" of the daily trading.[5] (Fig. 3)

[Place Figure 3 approximately here]

This compression allows treating daily imbalances as uniform images, which can be subjected to the deep learning algorithms.

Five randomly selected trading days (Oct. 7-8, 2019, Sept. 9, 2020, and Oct. 4-5, 2020) produced ten daily samples: one—with total market imbalance messages, the other—with ETF data only. We constructed from them five testing and five training samples according to the protocol exhibited in Fig. 4.

[Place Figure 4 approximately here]

A relative proportion of all-market and ETF-only samples according to the Fig. 3 is provided in Table 1.

---

[5] Each fingerprint contains 9216 pixels. We compress our ~400 MB daily database into ~200K text file, the compression of 2,000 times.

[Place Table 1 approximately here]

### 3. Metrics on the image space

Analysis of the output of neural networks is hard to rationalize. First, the human mind has been prepared by evolution to analyze two- or three-dimensional images in three-dimensional space. Tensor inputs, intermediate results, and outputs typical for the neural networks cannot be directly assessed by most humans. Second, the results of neural network analyses are necessarily stochastic and depend on the large number of estimated intrinsic parameters, which are frequently inaccessible, but in any case, too numerous to rationalize. Third, the results of deep learning can depend on the way the training and testing samples are being formed even from identical datasets. This can indicate the failure of the deep learning procedure (Brownlee, 2021) but they can also be indicative of additional information we fail to recognize.[6]

To inject some rationality into the results, we propose the following protocol. Namely, after the C-GAN generated fake images ("fingerprints") of the session, we consider these images as non-normalized probability distributions. Then we compute the pseudo-metric cosine between the image arrays $X$ and $Y$ according to the following formula:

$$C_{XY} = \frac{\|X+Y\|^2 - \|X-Y\|^2}{4\|X\|\cdot\|Y\|} \qquad (1)$$

In the above formula, the norm $\|.\|$ is a Frobenius norm of the matrix representing each of the image arrays. Because the number of epochs in a deep learning setting is variable and highly arbitrary, we compute the distance as the average of each twentieth of the last 400 images in the sequence. This is a first-stage averaging. Because, sometimes, the fake image comes out empty, i.e. with a zero norm, we modified this formula according to the following prescription:

$$C_{train,fake} = \frac{\|train+fake\|^2 - \|train-fake\|^2}{4\|train\|\cdot\|test\|}$$

$$C_{test,fake} = \frac{\|test+fake\|^2 - \|train-fake\|^2}{4\|train\|\cdot\|test\|} \qquad (2)$$

Equation (2) provides the answers close to the correct geometric formula (1) but does not fail in the case of an empty fake image. The pseudo-metric measure, calculated according to Equation (2) gives a fair picture of the affinity of the fake visual images to the originals (see Fig. 5) but it is still unstable with respect to different stochastic realizations of the fake images.

---

[6] The output from C-GAN indicates a deep learning failure called "mode collapse" by (Brownlee, 2021). Yet, the look-alike of the fake images remains excellent.

[Place Fig. 5 approximately here]

So, we apply a second stage averaging according to the formula for the mutual information:

$$MInfo = \frac{1}{N}\sum_{i=1}^{N} log_2 \left(\frac{C_{train,fake,i}}{C_{test,fake,i}}\right) \qquad (3)$$

In Equation (3), *N* is the number of independent runs of the network. Note, that this formula does not depend on whether we use a "geometrically correct" Equation (1), or computationally convenient Equation (2). Unlike separate $C_{train, fake}$, and $C_{test, fake}$ norms, which may vary widely between consecutive runs of the C-GAN, their ratio is pretty stable for a given training and testing samples. Furthermore, if one exchanges training and test samples, the argument of the summation of Equation (3) only change its sign. Theoretically, the sign change is exact, but numerically it is within the statistical limits.

Despite the complexity of the described procedure, the intuition behind it is quite simple. Convolutional Generative Adversarial Network generates many fake images inspired by a training sample. These images are compared with a test sample. If the fake images perfectly fit both the training and test samples, the mutual information is exactly zero. Positive mutual information or a higher distance between training and fake images than between the fake and test images means that the fake image was less informative than the training image. It implies that the affinity of training and test samples could have been coincidental. On the contrary, negative mutual information suggests that the fake approximates the test sample as well or is better off than the training sample and their affinity might be due to coincidence.

4. **The results of C-GAN analysis**

The results of the fingerprint analysis by the C-GAN are displayed in Tables 2 and 3. Each cell in Table 1 is the binary logarithm of the ratio of the distances of the generator-devised fingerprint between training and testing samples, respectively. From Table 2 we see that the mutual information generally decreases with diminishing fraction of market samples and increasing ETF samples. The fraction of the market vs. ETF samples in the testing samples demonstrate no visible tendency with one exception. When the testing file is almost entirely composed of the market samples, mutual information becomes exactly zero irrespective of the training samples.

[Place Tables 2 and 3 approximately here]

The column and the row averages are provided in Table 3. Testing sample 2 is an outlier. We tentatively attribute it to the fact that one of the day's data was, in some sense, exceptional. And indeed, the file for the 09-09-2020 contained, probably, only the stocks listed on NYSE, not all traded stocks. All averages are positive. This suggests that all-market files are easier to approximate by the generator-produced fakes than the ETF-only files. We consider it the evidence that ETF-only files have more distinguishing features than the all-market files and are, on average, more distant from fakes.

**Conclusion**

In this paper, I present a new econometric methodology to investigate the causality of the financial time series. My method was applied towards a solution to an important question: whether the liquidity of the markets is driven by the individual stocks or ETFs. As a measure of liquidity, I took the information content of the number of imbalances (a failure to complete a transaction) and the dollar value of the incomplete transactions. The information content was measured as a pseudo-distance between the time series in the two-dimensional state space (number of buckets and dollar imbalance buckets).

The preliminary answer is that both the rate of imbalance arrivals and the dollar value of resulting imbalances of the ETFs are more informative – in the sense of finer features non-reproducible by fakes – than the individual stocks with ETFs counted as separate stocks. This is not surprising empirically given the fact that the imbalance messages produced by 1,000+ ETFs constitute about one-eighth of the totality of exceptions in the entire database but the dollar value of their imbalances is about two-thirds of the entire dollar value of the market imbalance. It is not surprising theoretically because as it was pointed out (Nolan, 2016) (Evans, 2018), ETF securities are used for hedging much more frequently than individual stocks.

# Appendix

## Description of the TAQ ARCA messages

### 2.11 SYMBOL INDEX MAPPING MESSAGE (MSG TYPE '3')

| FIELD NAME | FIELD ORDER | FORMAT | DESCRIPTION |
|---|---|---|---|
| MsgType | 1 | Numeric | This field identifies the type of message. '3' – Sequence Number Reset message |
| SequenceNumber | 2 | Numeric | Message sequence number by channel. Must derive sequence number based on leading sequence number for each packet. |
| Symbol | 3 | NYSE Symbology | This field displays the symbol in NYSE symbology |
| SymbolIndex | 4 | Numeric | This field identifies the numerical representation of the NYSE symbol. This field is unique for products within each respective market and cannot be used to cross reference a security between markets. |
| Market ID | 5 | Numeric | ID of the Originating Market: <br> ■ '1' - NYSE Cash |

### 2.15 SECURITY STATUS MESSAGE (MSG TYPE '34')

| FIELD NAME | FIELD ORDER | FORMAT | DESCRIPTION |
|---|---|---|---|
| MsgType | 1 | Numeric | This field identifies the type of message. '34' – Security Status Message |
| SequenceNumber | 2 | Numeric | Message sequence number by channel. Must derive sequence number based on leading sequence number for each packet. |
| SourceTime | 3 | HH:MM:SS.nnnnnn | This field specifies the time when the msg was generated in the order book. The number represents the number of seconds at microsecond accuracy in UTC time (since EPOCH) |
| Symbol | 4 | NYSE Symbology | This field displays the symbol in NYSE symbology |
| SymbolSeqNum | 5 | Numeric | This field contains the symbol sequence number |
| Security Status | 6 | ASCII | The following are Halt Status Codes: <br> ■ '3' - Opening Delay |

## 2.9 IMBALANCE MESSAGE (MSG TYPE '105')

| FIELD NAME | FIELD ORDER | FORMAT | DESCRIPTION |
|---|---|---|---|
| Msg Type | 1 | Numeric | This field identifies the type of message. '105' – Imbalance Message |
| SequenceNumber | 2 | Numeric | Message sequence number by channel. Must derive sequence number based on leading sequence number for each packet. |
| SourceTime | 3 | HH:MM:SS.nnnnnn | This field specifies the time when the msg was generated in the order book. The number represents the number of seconds at microsecond accuracy in UTC time (since EPOCH) |
| Symbol | 4 | NYSE Symbology | This field displays the symbol in NYSE symbology |
| SymbolSeqNum | 5 | Numeric | This field contains the symbol sequence number |
| ReferencePrice | 6 | Numeric | The Reference Price is the Last Sale if the last sale is at or between the current best quote. Otherwise the Reference Price is the Bid Price if last sale is lower than Bid price, or the Offer price if last sale is higher than Offer price. |
| PairedQty | 7 | Numeric | This field contains the paired off quantity at the reference price point |
| TotalImbalanceQty | 8 | Numeric | This field contains the total imbalance quantity at the reference price point |
| MarketImbalanceQty | 9 | Numeric | This field indicates the total market order imbalance at the reference price |
| AuctionTime | 10 | hh:mm | Projected Auction Time |
| AuctionType | 11 | Alpha | ■ 'O' – Open (4am) Arca Only<br>■ 'M' – Market (9:30am)<br>■ 'H' - Halt<br>■ 'C' – Closing<br>■ 'R' – Regulatory Imbalance<br>**Note:** For the NYSE/MKT, the opening imbalance will have an "M" Auction Type |

# Message type '105'

| FIELD NAME | FIELD ORDER | FORMAT | DESCRIPTION |
|---|---|---|---|
| ImbalanceSide | 12 | Alpha | This field indicates the side of the imbalance Buy/sell. Valid Values:<br>■ 'B' – Buy<br>■ 'S' – Sell<br>■ ' ' – No imbalance<br>Note: This field is a future enhancement for NYSE Arca and will have a '0' value until such time. |
| ContinuousBook ClearingPrice | 13 | Numeric | The Continuous Book Clearing Price is defined as the price closest to last sale where imbalance is zero.<br>If a Book Clearing Price is not reached, the Clearing Price, a zero will be published in the Book Clearing Price Field<br>Note: This field is a future enhancement for NYSE Arca and will have a '0' value until such time. |
| ClosingOnly ClearingPrice | 14 | Numeric | This field contains the indicative price against closing only order only<br>Note: This field is a future enhancement for NYSE Arca and will have a '0' value until such time. |
| SSRFilingPrice | 15 | Numeric | This field contains the SSR Filing Price. This price is the price at which Sell Short interest will be filed in the matching in the event a Sell Short Restriction is in effect for the security.<br>Note: The SSR Filing price is based on the National Best Bid at 9:30am. This price remains static after the SSR Filing price has been determined.<br>Note: This field is a future enhancement for NYSE Arca and will have a '0' value until such time. |

## The format of TAQ ARCA messages with a nanosecond time stamp

34,10,00:24:58.796044288,CBO,1,P,~,,,,,,~,P

3,11,BANC,1,51,N,C,100,11.47,0,0,N,.0001,1

105,273294,09:29:34.061214976,UHT,33,66.21,170,1141,0, 0930,M,B,67.67,0,0,0,0,0,0,0,0, ,

**Table 1**. The proportion of the all-market and ETF-only samples in each simulation.

| Training file | Market:ETF | Test file | Market:ETF |
|---|---|---|---|
| tr1 | 100%:0% | tes1 | 0%:100% |
| tr2 | 80%:20% | tes2 | 20%:80% |
| tr3 | 60%:40% | tes3 | 40%:60% |
| tr4 | 40%:60% | tes4 | 60%:40% |
| tr5 | 20%:80% | tes5 | 80%:20% |

**Table 2**. The results of measuring MInfo (Equations 2-3) between the 20 fake images created by the generator, and the training and test images. C-GAN was run for 1600 epochs and the fake images were taken uniformly from the last 400 images.

|       | tr1 |        | tr2 |        | tr3 |        | tr4 |        | tr5 |        |
|-------|--------|---------|--------|---------|--------|---------|--------|---------|--------|---------|
| **tes1** | 0.7855 | 0.6538 | 0.1511 | 0.7556 | -0.4958 | -0.0203 | 0.2263 | -0.1734 | -0.6959 | 0.8837 |
| **tes2** | 0.5147 | -      | 1.0126 | -      | -0.2086 | -      | 0.5735 | -      | -1.4379 | -      |
| **tes3** | 2.5042 | 2.6258 | 2.2150 | 2.5771 | 2.2438  | 2.0920 | 2.5458 | 2.3841 | 2.0923  | 2.2976 |
| **tes4** | 0.3850 | 0.3785 | 0.4250 | 0.7784 | 0.7484  | 0.0000 | 0.1193 | 0.9425 | -0.0687 | -0.4286 |
| **tes5** | 0.0000 | -      | 0.0000 | -      | 0.0000  | -      | 0.0000 | -      | 0.0000  | -      |

**Table 3**. Mutual information (Equation 3) for the training and test samples

| Training file | Test file average | Test file | Training file average |
|---|---|---|---|
| tr1 | 0.9809 | tes1 | 0.2071 |
| tr2 | 0.9894 | tes2 | 0.0909 |
| tr3 | 0.5449 | tes3 | 2.3578 |
| tr4 | 0.8273 | tes4 | 0.3280 |
| tr5 | 0.3303 | tes5 | 0 |

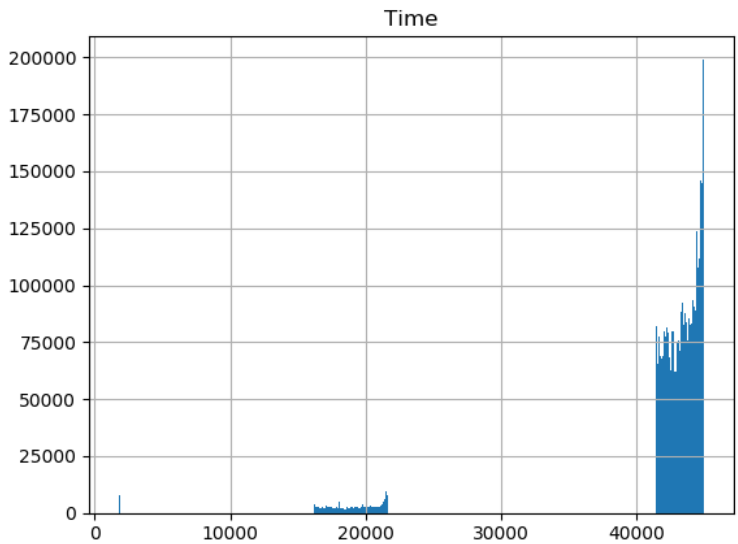

Fig. 1. Time density of the daily imbalance messages in the TAQ ARCA database. During 45,000 seconds of the daily operation of the system, there were around 4 million messages, the maximum coming at or around 4:00 PM. The maximum rate during a 100 second interval typically reached 200,000. As one can see most of the events are concentrated at the beginning, noontime, and the end-of-day.

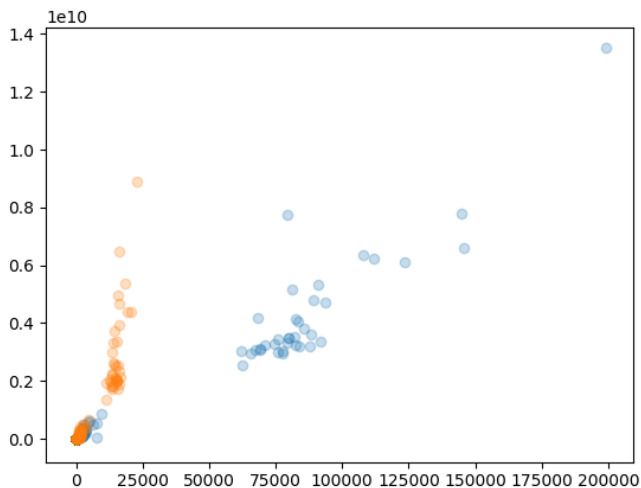

Fig. 2 Cumulative dollar imbalances as a function of the number of imbalance messages per 100 sec. periods. Orange dots are the transaction imbalances involving only the ETF shares and blue dots are the total market imbalance. We observe that the message rate of the ETFs is approximately one-eighth of the total market, but their dollar value exceeds 60% of the cumulative market imbalance.

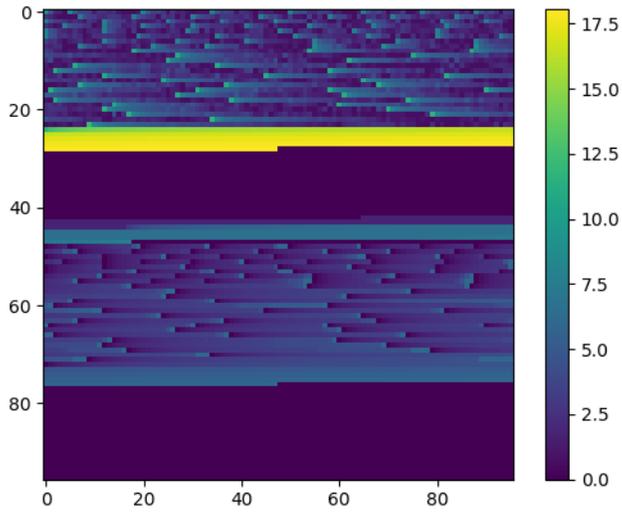

Fig. 3. Example of the "fingerprint" of a trading day on a natural logarithmic scale. The 105 message rate and the cumulative dollar amount of imbalances are placed into 96 bins.

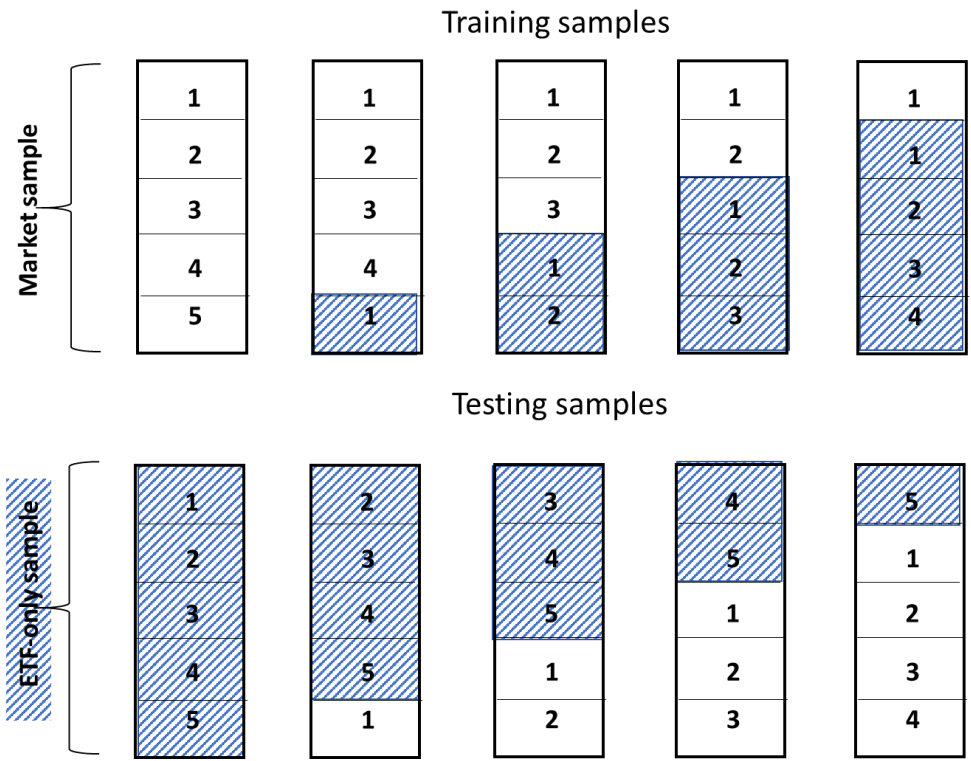

Fig. 4 Composition of the training and testing samples. Each number indicates a fingerprint of a given day in chronological order (10/7/2019, 10/8/2019, 09/09/2020, 10/4/2021, and 10/5/2021). Note, that only the first samples in corresponding rows are mirror images of each other.

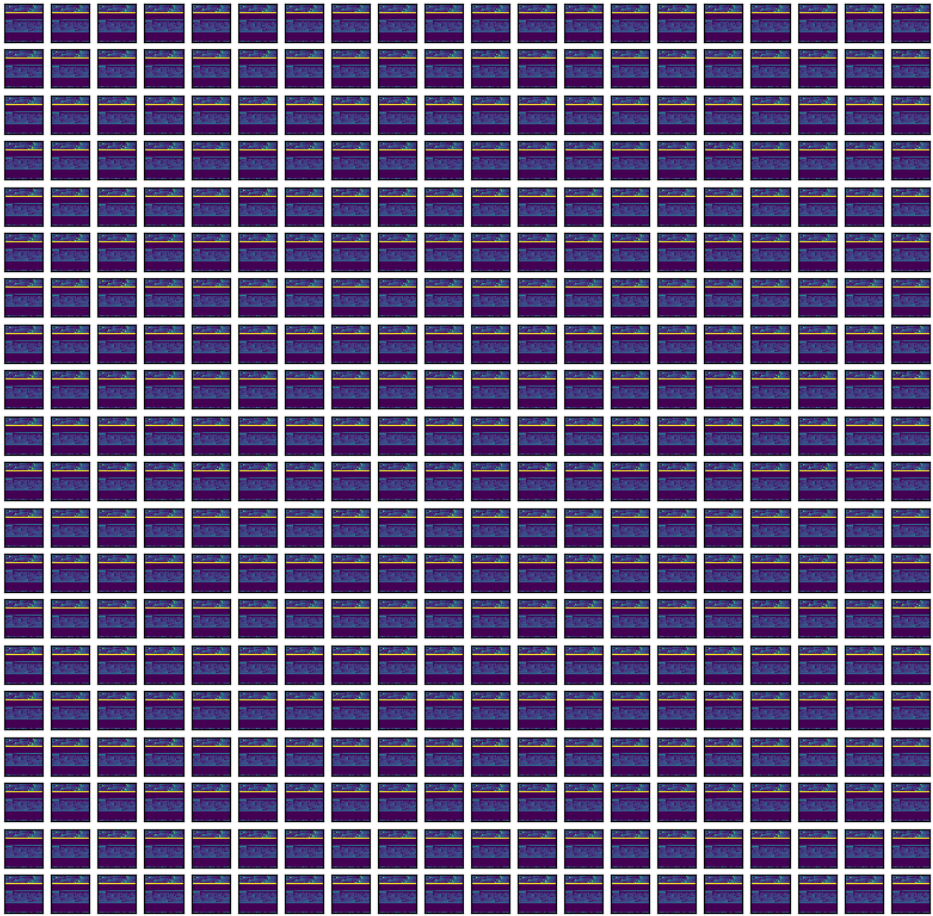

A)

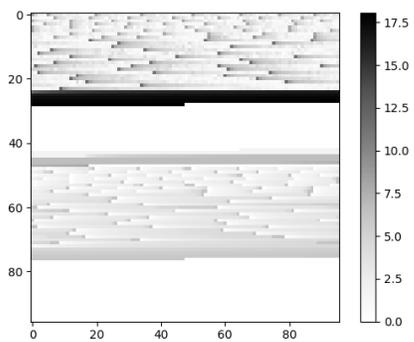

B)

Fig. 5 The output of C-GAN. A) Set of 400 fake images generated by the generator part of the deep learning network during 1600 epochs. B) An individual fake image (compare with Fig. 3).